\documentclass{ws-procs10x7}
\usepackage{balance}
\usepackage{multirow}

\newcolumntype{d}[1]{D{.}{.}{#1}}

\def\Journal#1#2#3#4{{\it #1} {\bf #2}, #3 (#4)}

\makeindex
\begin{document}

\title{Events with isolated leptons and missing transverse momentum at HERA}

\author{C. DIACONU$^*$ \\ on behalf of the H1 and ZEUS Collaborations}

\address{Centre de Physique des Particules de Marseille  and \\ Deutsches Elektronen Synchrotron, Notkestrasse 85, 22607 Hamburg, Germany\\$^*$E-mail: diaconu@cppm.in2p3.fr }

\twocolumn[\maketitle\abstract{Events with isolated leptons and missing transverse momentum are searched for in electron-- or positron--proton collisions at HERA by the H1 and ZEUS experiments. Such events may be explained within the Standard Model by  $W$ boson production, followed by leptonic decay. The analysis performed including HERA I and the recently collected data at HERA II results in clear evidence of $W$ production. However, H1 collaboration observes an excess of events in $e^+p$ collisions, not supported so far by ZEUS. If the difference between the H1 and ZEUS observations is attributed to a statistical fluctuation, the combined result exceeds the Standard Model prediction at $2.6\sigma$ level. A search for events with $\tau$ leptons and missing transverse momentum is also presented.  }
]

\section{Introduction}
HERA is an unique electron-- or positron--proton collider with a centre--of--mass energy of up to 320~GeV. The luminosity accumulated by each of the two collider--mode detectors, H1 and ZEUS, amounts to roughly 160 pb$^{-1}$ in $e^+p$ collisions and close to 200~pb$^{-1}$ in $e^-p$. This data sample enables the search for rare phenomena, with cross sections around or below 1~pb.
One such process is the production of $W$ bosons, for which the total production cross section is around 1.3pb$^{-1}$, calculated including NLO-QCD corrections~\cite{Diener:2002if}. If the $W$ boson decays leptonically, the corresponding events contain an energetic, isolated lepton and significant missing energy due to the escaping neutrino.
\par
Such events have been observed at HERA by the H1 collaboration~\cite{h1_hera1}. Moreover, an excess of events with large hadronic transverse momentum $P_T^X$ was reported after the first data taking period HERA I (1994--2000, $118$~pb$^{-1}$), where 11 events are observed with $P_T^X>25$~GeV for a Standard Model (SM) expectation of $3.5\pm 0.6$. The ZEUS collaboration also performed a search for this event topology, within an analysis aimed at a search for anomalous top production~\cite{zeus_hera1}, but did not confirm the excess observed by H1.
In this paper, the results from the H1 analysis~\cite{h1conf} performed including all available data up to July 2006 ($341$~pb$^{-1}$) and a new ZEUS analysis~\cite{zeusconf} using a data sample corresponding to an integrated luminosity of  $249$~pb$^{-1}$ are presented.
\section{Event selection and results}
Events with isolated leptons and missing transverse momentum are selected using the criteria  summarised in table~\ref{tabsel}. 
\begin{table}
\tbl{Main selection requirements in the H1 and ZEUS 
analyses.
 \label{tabsel}}
{\begin{tabular}{@{}l|ll@{}}
              & \multicolumn{1}{c}{H1}   & \multicolumn{1}{c}{ZEUS} \\ \hline
              & & \\
              & $P_T^e>10$~GeV  & $P_T^e>10$~GeV  \\
              & $P_T^{miss}>12$~GeV & $P_T^{miss}>12$~GeV \\
              &$5^\circ < \theta_e < 140^\circ$& $17^\circ < \theta_e < 86^\circ$\\
\begin{rotate}{90}{$e+P_T^{miss}$}{}\end{rotate} & -- & $P_T^{X}>12$~GeV \\
\hline
 & & \\
 & $P_T^\mu>10$~GeV & $P_T^\mu>10$~GeV  \\
 & $P_T^{miss}>12$~GeV & $P_T^{miss}>12$~GeV\\
 &$P_T^{X}>12$~GeV &$P_T^{X}>12~GeV$  \\
\begin{rotate}{90}{$\mu+P_T^{miss}$}{}\end{rotate}  &$5^\circ < \theta_\mu < 140^\circ$ & $17^\circ < \theta_\mu < 115^\circ$ \\

\end{tabular}}
\end{table}
\begin{table*}[t]
\tbl{The observed events and the SM expectation in the H1 and ZEUS analyses.  \label{tab}}
{\begin{tabular}{@{}ll|ccc@{}}
    \multicolumn{2}{l|}{\bf  $\mathrm{e^\pm p}$ Data  Preliminary} &
    Electron &
    Muon &
    Combined \\
    \multicolumn{1}{c}{H1} & \multicolumn{1}{l|}{1994-2006 341 pb$^{-1}$} &
    obs./exp. &
    obs./exp. &
    obs./exp. \\
    \multicolumn{1}{c}{ZEUS} & \multicolumn{1}{l|}{1998-2005 249 pb$^{-1}$ }&
 & & \\
    \hline
    {\footnotesize H1} &
    {\footnotesize $e$:Full Sample/$\mu:P_{T}^{X}~>12$~GeV } &
    {\footnotesize 35 / 34.0 $\pm$ 4.7 }&
    {\footnotesize 11 /  9.0 $\pm$ 1.4 }&
    {\footnotesize 46 / 43.0 $\pm$ 6.0 }\\
    {\footnotesize ZEUS} &
    {\footnotesize $P_{T}^{X}~>12$~GeV} &
    {\footnotesize  9 / 7.8 $\pm$ 0.6 }&
    {\footnotesize  6 /  5.9 $\pm$ 0.4}&
    {\footnotesize 15 / 13.7 $\pm$ 0.7 }\\
    \hline
    {\footnotesize H1} &
    {\footnotesize $P_{T}^{X}~>25$~GeV} 
&
    {\footnotesize 12 /  6.1 $\pm$ 1.1 }&
    {\footnotesize  6 /  5.4 $\pm$ 0.9 }& 
    {\bf \footnotesize 18 / 11.5 $\pm$ 1.8 }\\
    {\footnotesize ZEUS} &
    {\footnotesize $P_{T}^{X}~>25$~GeV} &

    {\footnotesize 4 /  4.4 $\pm$ 0.5 }&
    {\footnotesize  3 /  3.1 $\pm$ 0.3 }& 
    {\bf \footnotesize 7 / 7.5 $\pm$ 0.6 }\\
  \end{tabular}}
\end{table*}
\begin{figure*}[ht]
\centerline{
\epsfxsize=6.8cm\epsfysize=6.4cm\epsfbox{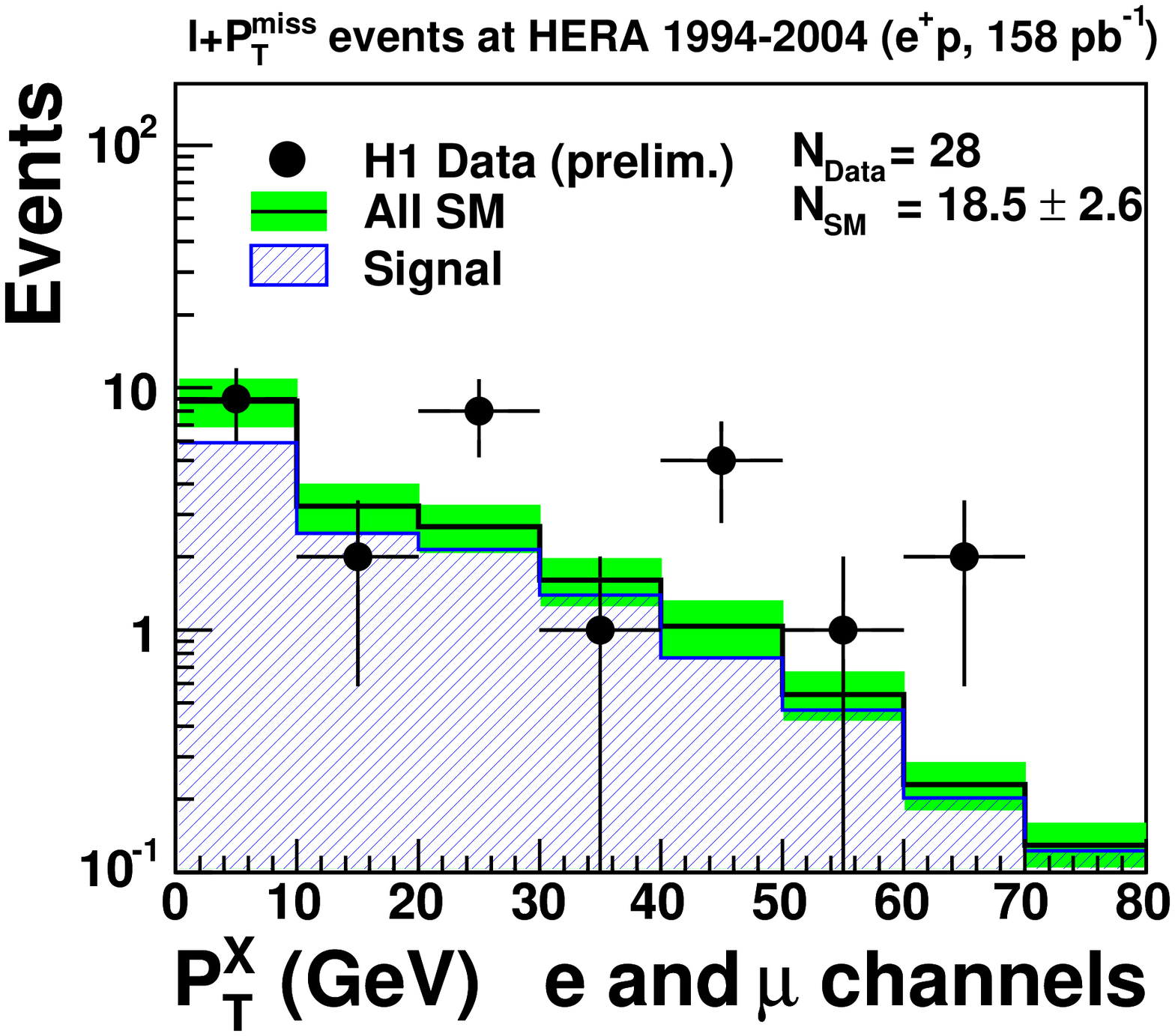}
\epsfxsize=6.8cm\epsfysize=6.4cm\epsfbox{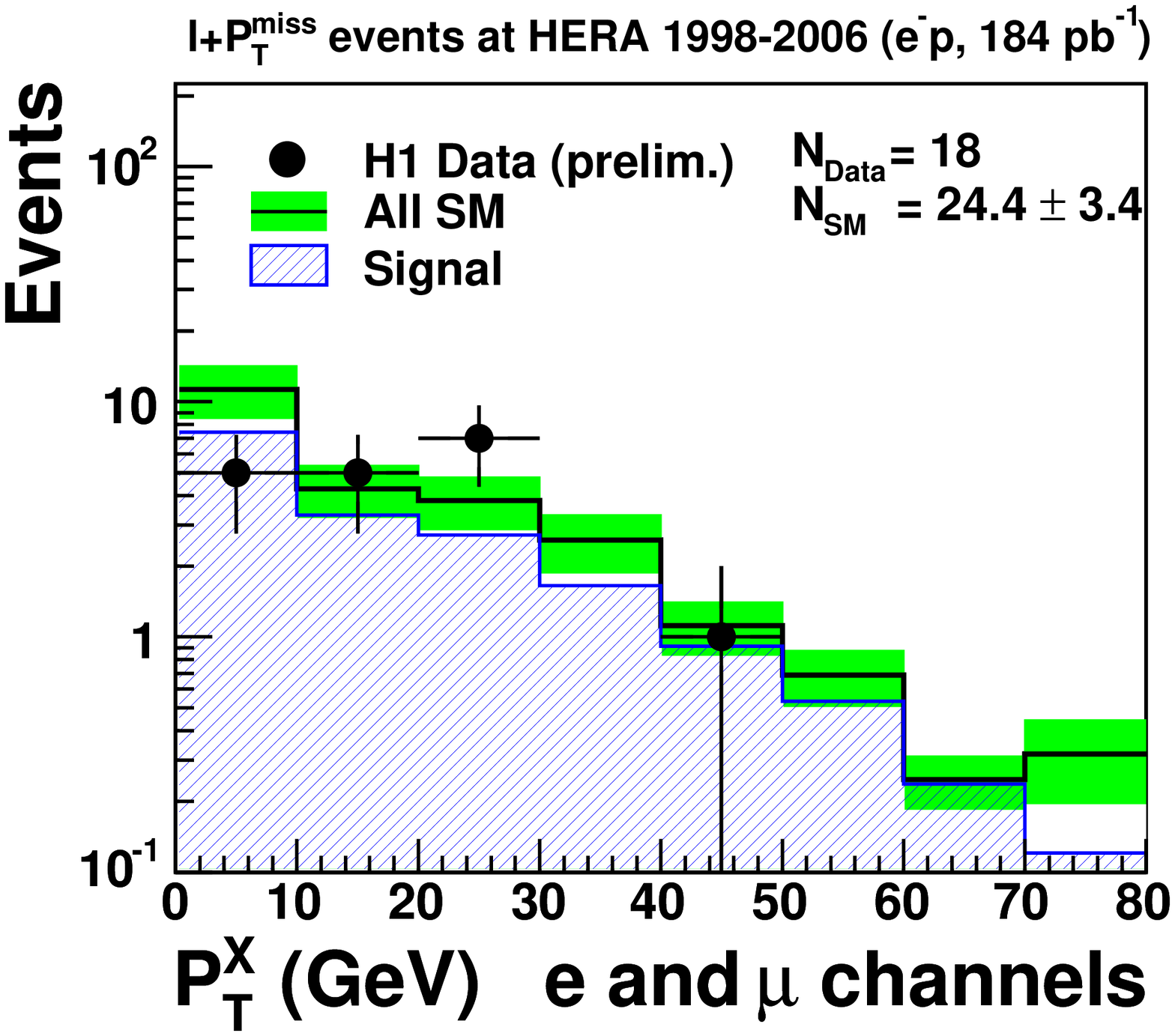}
}
\caption{The distributions of the observed events as a function of $P_T^X$ in the H1 analysis in $e^+p$ data (left) and $e^-p$ data (right). \label{fig1}}
\end{figure*}
\begin{table*}[t]
\tbl{The observed and expected numbers of events in the region $P_T^X>25$~GeV in H1 and ZEUS analyses.\label{tab25}}
{\begin{tabular}{@{}lll|ccc@{}}
    \multicolumn{3}{l|}{ \bf  $\mathrm{e^\pm p}$ Data  Preliminary} &
    Electron &
    Muon &
    Combined \\
    \multicolumn{2}{c}{ \bf $P_T^X>25$~GeV} & \multicolumn{1}{l|}{} &
    obs./exp. &
    obs./exp. &
    obs./exp. \\
    \multicolumn{2}{c}{} & \multicolumn{1}{l|}{}&
 & & \\
    \hline
  & {\footnotesize  H1 } &
    {\footnotesize  158~pb$^{-1}$} &
    {\footnotesize  9 / 2.3 $\pm$ 0.4 }&
    {\footnotesize  6 / 2.3 $\pm$ 0.4 }&
    {\footnotesize 15 / 4.6 $\pm$ 0.8 }\\
    
 &  {\footnotesize  ZEUS } &
    {\footnotesize  106~pb$^{-1}$} &
    {\footnotesize 1 / 1.5 $\pm$ 0.1 }&
    {\footnotesize 1 / 1.5 $\pm$ 0.2 }&
    {\footnotesize 2 / 3.0 $\pm$ 0.3 }\\
    
\begin{rotate}{90}{\boldmath{$\; e^+p$}}{}\end{rotate} &
    {\footnotesize  H1+ZEUS } &
    {\footnotesize  264~pb$^{-1}$} &
    {\footnotesize 10 / 3.8 $\pm$ 0.5 }&
    {\footnotesize 7 / 3.8 $\pm$ 0.6 }&
    {\footnotesize \bf 17 / 7.6 $\pm$ 1.1 }\\
\hline
  &  {\footnotesize H1 } &
    {\footnotesize  184~pb$^{-1}$} &
    {\footnotesize 3 / 3.8 $\pm$ 0.6 }&
    {\footnotesize 0 / 3.1 $\pm$ 0.5 }&
    {\footnotesize 3 / 6.9 $\pm$ 1.1 }\\
    
 &   {\footnotesize ZEUS } &
    {\footnotesize  158~pb$^{-1}$} &
    {\footnotesize 3 / 2.9 $\pm$ 0.5 }&
    {\footnotesize 2 / 1.6 $\pm$ 0.2 }&
    {\footnotesize 5 / 4.5 $\pm$ 0.7 }\\
    
\begin{rotate}{90}{\boldmath{$\; e^-p$}}{}\end{rotate} &
    {\footnotesize H1+ZEUS } &
    {\footnotesize  327~pb$^{-1}$} &
    {\footnotesize 6 / 6.7 $\pm$ 1.1 }&
    {\footnotesize 2 / 4.7 $\pm$ 0.7 }&
    {\footnotesize \bf 8 / 11.4 $\pm$ 1.8 }\\
  \end{tabular}}
\end{table*}
\begin{figure*}[ht]
\centerline{
\epsfxsize=7.2cm\epsfysize=6.7cm\epsfbox{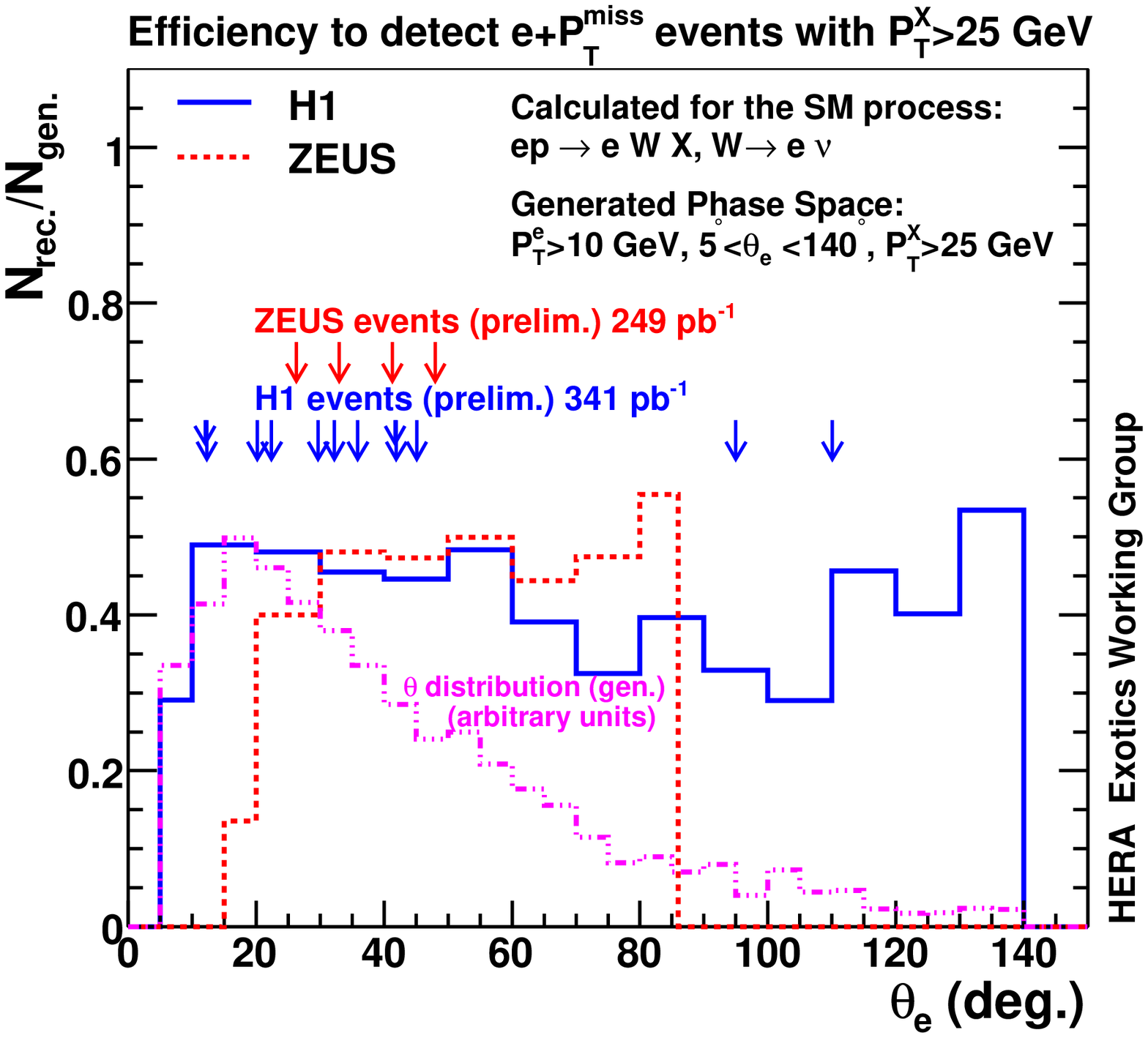}
\epsfxsize=7.2cm\epsfysize=6.7cm\epsfbox{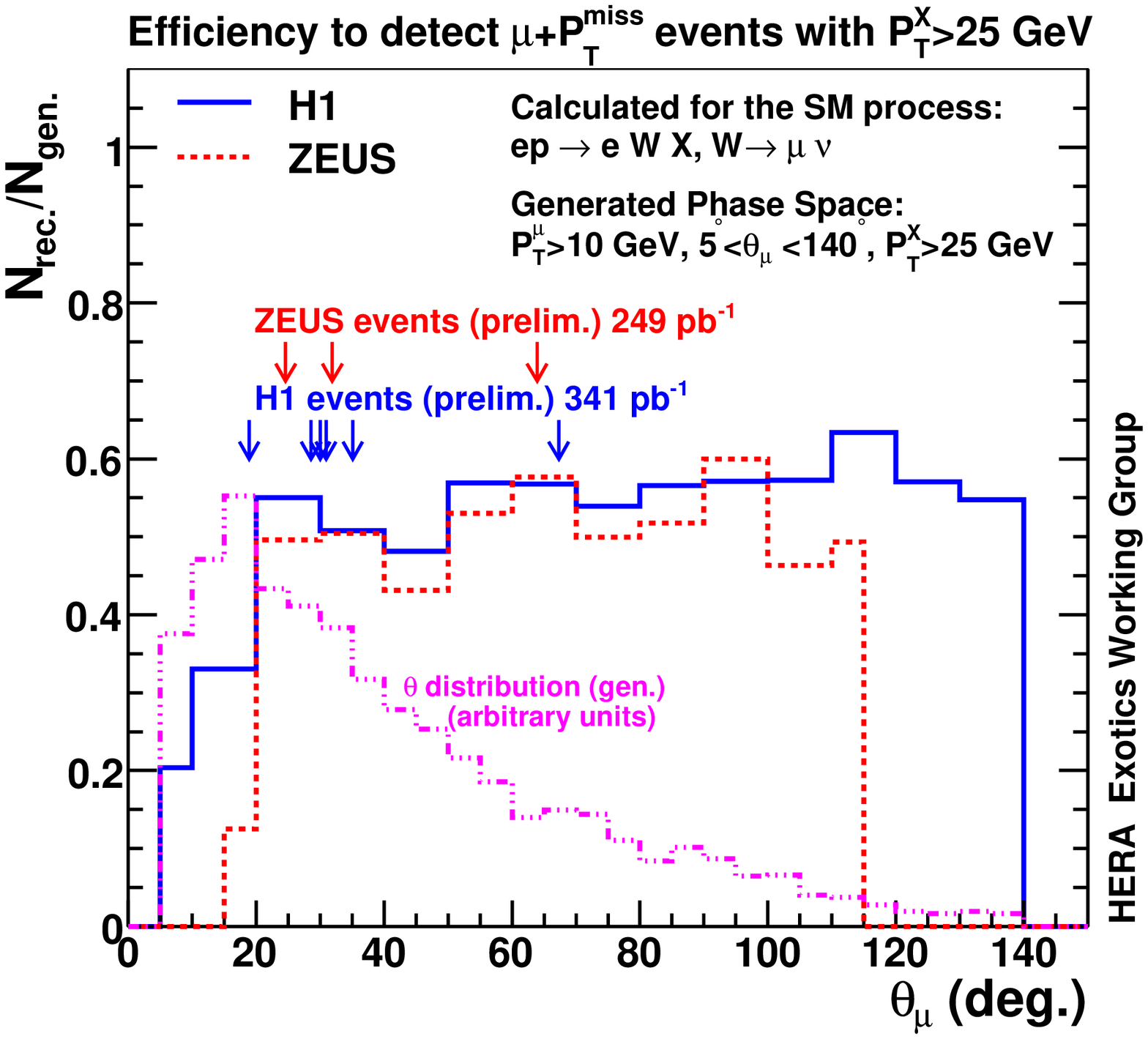}
}
\caption{The acceptances of the H1 and ZEUS $e+P_T^{miss}$ (left) and $\mu+P_T^{miss}$ (right) analyses as a function of the lepton polar angle.  \label{fig2}}
\end{figure*}

The lower thresholds for the missing and lepton tranverse momenta are identical in both analyses. Main differences are related to the conditions -- more restrictive for ZEUS analysis -- applied to the lepton polar angle (defined w.r.t. proton beam direction) and to the hadronic transverse momentum $P_T^X$.
Extra background suppression criteria, discussed elsewhere~\cite{h1conf,zeusconf}, are applied by both analyses to further suppress background processes 
that 
may enter the $\ell+P_T^{miss}$ selection only due to measurement fluctuations leading to fake leptons or fake $P_T^{miss}$.
\par
The results obtained after these selections are shown in table~\ref{tab}. A good agreement is observed with the SM predictions in the full phase space of the analysis. The observed number of events in the electron channel in H1 analysis is significantly higher than in the other cases since it covers the region $P_T^X<12$~GeV, where the signal is expected to dominate. The purity in signal (dominated by the $W$ production) is of $60$--$68$\% in the electron channel and $76$--$83$\% in the muon channel. This result provides therefore clear evidence of single $W$ boson production at HERA. 
At large hadronic transverse momentum $P_T^X>25$~GeV, a number of events in excess w.r.t. the SM prediction is observed by the H1 Collaboration, while good agreement is seen in the ZEUS analysis. 
\par
The distribution of events in the  H1 analysis as a function of $P_T^X$ is shown separately in $e^+p$ and $e^-p$ data samples in figure~\ref{fig1}.  This result indicates that the excess of events at large $P_T^X$ originates from the $e^+p$ data sample. The observations of H1 and ZEUS analyses at $P_T^X>25$~GeV are shown in table~\ref{tab25}. The excess observed by H1 in $e^+p$ data has a significance of 3.4~$\sigma$ but is not confirmed by the ZEUS analysis. If the difference between H1 and ZEUS observations is attributed to a statistical fluctuation, the results can be combined and the total observed yield is  $17$ events compared to $7.6\pm 1.1$ expected from the SM, corresponding to a $2.6$~$\sigma$ deviation. 
In the $e^-p$ data sample, a good agreement with the SM is observed by both H1 and ZEUS, with a total of $8$ events observed for $11.4\pm 1.8$ expected from the SM. 

\section{Comparison between H1 and ZEUS analyses}
The different observations of H1 and ZEUS at large $P_T^X$ have been investigated using a sample of Monte Carlo 
events with $W$ decays to electrons or muons, reconstructed and analysed in the H1 and ZEUS frameworks, respectively. The efficiency, defined as the ratio of the reconstructed to the generated number of events in each bin, is shown as a function of the polar angle of the lepton in figure~\ref{fig2}.  The main difference between the polar angle acceptance domain between the H1 and ZEUS analyses is visible. In the ZEUS analysis domain, the H1 analysis has similar acceptance. The majority of the events with large $P_T^X$ observed by H1 are within the ZEUS acceptance. The H1 analysis extends more in the  forward direction, where the fraction of a possible signal can be significant. This is illustrated by  the distribution of the polar angle of the charged lepton produced in $W$ production process, also shown in figure~\ref{fig2}. This behaviour is typical also for processes beyond the SM, involving production of particles with large masses, for which partons with large momentum  collide with the electrons, giving rise to a large boost of the produced particles in the forward direction.

\section{Search for events with isolated $\tau$ leptons and missing transverse momentum}
To complement the observation reported above, the H1 and ZEUS collaborations have searched for events containing isolated tau leptons and missing transverse momentum. In each event,  the tau lepton candidate decay into hadrons is identified as a narrow, low multiplicity jet. Both H1 and ZEUS analyses search for one--prong hadronic tau decays by requiring one isolated charged track associated with a narrow calorimetric deposit. The H1 analysis~\cite{tauh1} uses a simple cut on the jet radius $R_{\mathrm jet}<0.12$, calculated as the energy weighted average distance in pseudo-rapidity to the jet axis of the clusters composing the jet.
The ZEUS analysis~\cite{tauzeus} combines several shape variables calculated from the calorimetric deposit in a discriminant used to disentangle between signal and background and requires an additional jet with $P_T^{\mathrm jet}>5$~GeV in each event.  
\begin{table}
\tbl{Results of searches for events with $\tau$ leptons and missing transverse momentum. The ZEUS analysis requires a jet with $P_T^{\mathrm jet}>5$~GeV and is based mostly on data collected in $e^+p$ collisions}
{\begin{tabular}{l|ll@{}}
$\tau+P_T^{\mathrm miss}$ & \footnotesize  All & \footnotesize  $P_T^X>25$~GeV \\ 
\footnotesize  Analysis & & \\ \hline
\footnotesize  H1 $e^+p$ 158~pb$^{-1}$ &{\footnotesize $8/10.6\pm2.9$} & {\footnotesize $0/0.40\pm0.10$} \\ 
\footnotesize H1 $e^-p$ 121~pb$^{-1}$&{\footnotesize $17/13.5\pm2.6$} & {\footnotesize $3/0.40\pm0.12$} \\
\footnotesize ZEUS 130~pb$^{-1}$& {\footnotesize  $3/0.40\pm0.12$} & {\footnotesize $2/0.20\pm0.05$} \\ \hline
\end{tabular}}
\label{tabtau}
\end{table}
\par 
The results are summarised in table~\ref{tabtau}. The ZEUS analysis is based on HERA I data, collected mostly in positron--proton collisions. The H1 preliminary analysis extends to the HERA II data sample and is shown separately for electron-- and positron--proton collisions. A few candidates are observed at large hadronic transverse momentum, in agreement with the SM prediction.
\par The rate observed in the full analysis phase space is lower in the ZEUS analysis due to the jet requirement described above, which reduces both the signal from $W$ production and the background, dominated by the charged current (CC) process. The expected signal rate is much lower than the one expected in the $e+P_T^{miss}$ and $\mu+P_T^{miss}$ analyses, due to lower efficiency to identify tau decays. In addition, the purity is lower due to irreducible background from CC events with hadronic jets fluctuating to low multiplicity.

\section*{Acknowledgments}
I would like to thank my colleagues from H1 and ZEUS Collaborations and in particular Emmanuelle Perez, Max Klein, Elisabetta Gallo, Gerhard Brandt, Dave South, James Ferrando and Katherine Korcsak-Gorzo for kind support and fruitful common work.

\balance

\end{document}